\documentstyle[12pt]{article}

\def\bseq{\begin{subequation}}  
\def\eseq{\end{subequation}}
\def\bsea{\begin{subeqnarray}}  
\def\esea{\end{subeqnarray}}

\def\beq{\begin{equation}}
\def\eeq{\end{equation}}
\def\eea{\end{eqnarray}}
\def\bq{\begin{quote}}
\def\eq{\end{quote}}

\newcommand{\EQ}{\begin{equation}}
\newcommand{\EN}{\end{equation}}
\newcommand{\bea}{\begin{eqnarray}}
\newcommand{\ena}{\end{eqnarray}}

\renewcommand{\a}{\alpha}
\renewcommand{\b}{\beta}

\renewcommand{\d}{\delta}
\newcommand{\th}{\theta}

\newcommand{\pa}{\partial}

\newcommand{\G}{\Gamma}

\newcommand{\e}{\epsilon}

\newcommand{\k}{\kappa}

\renewcommand{\L}{\Lambda}
\newcommand{\m}{\mu}

\newcommand{\p}{\pi}

\newcommand{\s}{\sigma}

 \newcommand{\Db}{\bar{D}}

\def\Mb{\kern 2pt\mathchoice
            {
                 \vbox{\hrule width10pt height 0.4pt depth 0pt
                 \kern 1.2pt\hbox{\kern -2pt$\displaystyle M$}}}
            {
                 \vbox{\hrule width10pt height 0.4pt depth 0pt
                 \kern 1.2pt\hbox{\kern -2pt$\textstyle M$}}}
            {
                 \vbox{\hrule width6pt height 0.4pt depth 0pt
                 \kern 1.0pt\hbox{\kern -2pt$\scriptstyle M$}}}
            {
                 \vbox{\hrule width5pt height 0.4pt depth 0pt
                 \kern 0.8pt\hbox{\kern -2pt$\scriptscriptstyle M$}}}}

\def\Sb{\kern 2pt\mathchoice
            {
                 \vbox{\hrule width6pt height 0.4pt depth 0pt
                 \kern 1.2pt\hbox{\kern -2pt$\displaystyle S$}}}
            {
                 \vbox{\hrule width6pt height 0.4pt depth 0pt
                 \kern 1.2pt\hbox{\kern -2pt$\textstyle S$}}}
            {
                 \vbox{\hrule width3.5pt height 0.4pt depth 0pt
                 \kern 1.0pt\hbox{\kern -2pt$\scriptstyle S$}}}
            {
                 \vbox{\hrule width3pt height 0.4pt depth 0pt
                 \kern 0.8pt\hbox{\kern -2pt$\scriptscriptstyle S$}}}}

\def\Rb{\kern 2pt\mathchoice
            {
                 \vbox{\hrule width5.5pt height 0.4pt depth 0pt
                 \kern 1.2pt\hbox{\kern -2.5pt$\displaystyle R$}}}
            {
                 \vbox{\hrule width5.5pt height 0.4pt depth 0pt
                 \kern 1.2pt\hbox{\kern -2.5pt$\textstyle R$}}}
            {
                 \vbox{\hrule width3.5pt height 0.4pt depth 0pt
                 \kern 1.0pt\hbox{\kern -2.2pt$\scriptstyle R$}}}
            {
                 \vbox{\hrule width3pt height 0.4pt depth 0pt
                 \kern 0.8pt\hbox{\kern -2.2pt$\scriptscriptstyle R$}}}}

  \def\pp{{\mathchoice
          {
              \kern 1pt%
              \raise 1pt
              \vbox{\hrule width5pt height0.4pt depth0pt
                    \kern -2pt
                    \hbox{\kern 2.3pt
                          \vrule width0.4pt height6pt depth0pt
                          }
                    \kern -2pt
                    \hrule width5pt height0.4pt depth0pt}%
                    \kern 1pt
           }
            {
              \kern 1pt%
              \raise 1pt
              \vbox{\hrule width4.3pt height0.4pt depth0pt
                    \kern -1.8pt
                    \hbox{\kern 1.95pt
                          \vrule width0.4pt height5.4pt depth0pt
                          }
                    \kern -1.8pt
                    \hrule width4.3pt height0.4pt depth0pt}%
                    \kern 1pt
            }
            {
              \kern 0.5pt%
              \raise 1pt
              \vbox{\hrule width4.0pt height0.3pt depth0pt
                    \kern -1.9pt  
                    \hbox{\kern 1.85pt
                          \vrule width0.3pt height5.7pt depth0pt
                          }
                    \kern -1.9pt
                    \hrule width4.0pt height0.3pt depth0pt}%
                    \kern 0.5pt
            }
            {
              \kern 0.5pt%
              \raise 1pt
              \vbox{\hrule width3.6pt height0.3pt depth0pt
                    \kern -1.5pt
                    \hbox{\kern 1.65pt
                          \vrule width0.3pt height4.5pt depth0pt
                          }
                    \kern -1.5pt
                    \hrule width3.6pt height0.3pt depth0pt}%
                    \kern 0.5pt
            }
        }}

  \def\mm{{\mathchoice
                       {
                             \kern 1pt
               \raise 1pt    \vbox{\hrule width5pt height0.4pt depth0pt
                                  \kern 2pt
                                  \hrule width5pt height0.4pt depth0pt}
                             \kern 1pt}
                       {
                            \kern 1pt
               \raise 1pt \vbox{\hrule width4.3pt height0.4pt depth0pt
                                  \kern 1.8pt
                                  \hrule width4.3pt height0.4pt depth0pt}
                             \kern 1pt}
                       {
                            \kern 0.5pt
               \raise 1pt
                            \vbox{\hrule width4.0pt height0.3pt depth0pt
                                  \kern 1.9pt
                                  \hrule width4.0pt height0.3pt depth0pt}
                            \kern 1pt}
                       {
                           \kern 0.5pt
             \raise 1pt  \vbox{\hrule width3.6pt height0.3pt depth0pt
                                  \kern 1.5pt
                                  \hrule width3.6pt height0.3pt depth0pt}
                           \kern 0.5pt}
                       }}

\def\pd{{\kern0.5pt
                   + \kern-5.05pt \raise5.8pt\hbox{$\textstyle.$}\kern 0.5pt}}

\def\pmd{{\kern0.5pt
                  \pm \kern-5.05pt \raise6.3pt\hbox{$\textstyle.$}\kern1.5pt}}

\def\md{\mathchoice
   {
      {{\kern 1pt - \kern-6.2pt \raise5pt\hbox{$\textstyle.$}\kern 1pt}}}
    {
      {{\kern 1pt - \kern-6.2pt \raise5pt\hbox{$\textstyle.$}\kern 1pt}}}
    {
      {\kern0.5pt - \kern-5.05pt \raise3.4pt\hbox{$\textstyle.$}\kern0.5pt}}
    {
      {\kern0.5pt - \kern-5.05pt \raise3.4pt\hbox{$\textstyle.$}\kern0.5pt}}}

\newcommand{\Del}{\nabla}

\renewcommand{\thefootnote}{\fnsymbol{footnote}}

\begin{document}

\newpage
\begin{titlepage}
\begin{flushright}
{IFUM-492-FT}\\
{BRX-TH-369}
\end{flushright}
\vspace{2cm}
\begin{center}
{\bf {\large SUPERGRAVITY DRESSING OF $\b$-FUNCTIONS IN N=1 AND N=2
SUPERSYMMETRIC MODELS}}\\
\vspace{1.5cm}
M. T. Grisaru
\\
\vspace{1mm}
{\em Physics Department, Brandeis University, Waltham, MA 02254, USA}\\
\vspace{2mm}
and\\
\vspace{2mm}
D. Zanon\\
\vspace{1mm}
{\em  Dipartimento di Fisica dell'Universit$\grave{a}$ di Milano and INFN,
Sezione di Milano, Via Celoria 16, I-20133 Milano, Italy}

\vspace{1.1cm}
{{ABSTRACT}}
\end{center}

\bq
We discuss the dressing of one-loop  $\s$-model $\b$-functions by induced
supergravity,
for both $N=1$ and $N=2$ supersymmetric theories. We obtain exact results by a
 superconformal gauge argument, and verify them in the semi-classical
limit  by
explicit perturbative calculations in the light-cone gauge. We find that for
$N=2$
theories there is no dressing of the one-loop $\b$-functions.
\eq

\vfill

\begin{flushleft}
Feb. 1995

\end{flushleft}
\end{titlepage}

\newpage

\renewcommand{\thefootnote}{\arabic{footnote}}
\setcounter{footnote}{0}
\newpage
\pagenumbering{arabic}


Recently, a number of papers have appeared dealing with the
gravitational dressing of  one-loop $\b$-functions for two-dimensional systems
away from their
conformal point  \cite{Schmid,Kogan,Ambjorn,Tanii}. Although using different
approaches, they conclude that the effect of induced gravity \cite{Polyakov1}
is to rescale the
one-loop $\b$-functions by an overall factor
\EQ
\b_G = \frac{\k+2}{\k+1} \b
\EN
\EQ
\k+2= \frac{1}{12}[c- 13-\sqrt{(1-c)(25-c)}]
\EN
where $\k$ is the central charge of the gravitational $SL(2R)$ current algebra.
An alternative way of writing this result is
\EQ
\b_G =-\frac{2}{\a_{_+} Q} \b
\EN
where
\bea
Q&=&\sqrt{\frac{25-c}{3}} \nonumber\\
\a_{_+} &=& \frac{1}{2}\left[-Q+\sqrt{Q^2-8}\right]
\ena
This result
appears to be universal, holding for conformal field theories
perturbed by some marginal operator, or for two-dimensional $\s$-models
away from their fixed point. Beyond one loop the situation is less clear,
although indications exist that the dressing may not be universal \cite{Dorn}.

The one-loop result has been obtained by presenting induced gravity in
conformal
gauge, i.e. in its Liouville incarnation \cite{Schmid,Ambjorn}, in light-cone
gauge
using the nonlocal form of the induced action \cite{Kogan},  and by studying
the
problem in $2+\e$ dimensions \cite{Tanii}. The first method  attributes the
effect to  the difference
between the scale defined by the fiducial metric, and that  defined by the
physical
metric determined by the Liouville mode. The second method relies on the
induced
gravity Ward identities \cite{Polyakov2} and the corresponding dressing of
correlation functions. The third method is similar to the first, insofar as it
relies on the presence of the (induced)  renormalized cosmological constant.

In this work we extend the above results to the cases of  $(1,1)$ and $(2,2)$
supergravity. We present a general argument, similar in spirit to that of
the above references, relying on the general KPZ \cite{KPZ} and DDK
\cite{David,Distler,DHK}
results, and we supplement it with perturbative verifications.

The general argument  in (super)conformal gauge, for ordinary, $N=0$, gravity,
and for $N=1$ or $N=2$ supergravity  is based on  the following
idea: the dressed $\b$-functions are defined by the response of systems
to changes in the physical scale which in the presence of  the Liouville field
gets modified
with respect to the standard renormalization scale. In two-dimensions, a
sensible definition of the physical scale is provided by the cosmological
constant,
the only dimensionful  object in the theory. In light-cone gauge the
cosmological
term is just a $c$-number, so that the usual scaling is physical and
modifications
of the matter $\b$-functions arise through new divergent contributions due to
the
gravitational couplings. In conformal gauge instead, the one-loop matter
divergence does not receive gravitational corrections. However, in this case,
for
$N=0$ and $N=1$ theories the cosmological constant is renormalized
by quantum corrections:
\EQ
\L_0 \int d^2z d^{2N}\theta~  e^{\phi} \rightarrow
\L_R \int d^2z d^{2N}\theta :e^{\a_{_+} \phi} :
\EN
where
\EQ
\L_0=\m^{2s} \L_R Z
\EN
In the above relations $\m$ is the renormalization mass,
$Z=(\m^2 a^2)^{\frac{\a_{_+} ^2}{2}}$, and $\a_{_+}$ is the positive root of
\EQ
- \frac{1}{2} \a (\a +Q) =s
\EN
i.e.
\EQ
\a_{_+}=\frac{1}{2} \left[  -Q +\sqrt{Q^2-8s }\right]
\EN
For the $N=0,1$ theories one has
\bea
N&=&0: ~~~s=1 ~~~~~~~Q=\sqrt{\frac{25-c}{3}} \nonumber\\
N&=&1: ~~~s=\frac{1}{2}~~~~~~~Q= \sqrt{\frac{9-c}{2}}
\ena

{From} Eq. (6) we obtain
\EQ
\frac{\pa ln\L_R}{\pa ln\m} = -(2s+\a_{_+}^2) =\a_{_+} Q
\EN
Therefore
\EQ
\b_G=\frac{\pa ln\m}{\pa ln(\L_R)^{-\frac{1}{2s}}} \b=
-\frac{2s}{\a_{_+} Q} \b
\EN
Using the above expressions, one finds for the ordinary gravity case, $N=0$,
the
result in Eq. (1).
 For $N=1$, using also  the expression for the level $\k$ of the light-cone
supergravity Ka\v{c}-Moody algebra
\bea
N=1:~~~~\k+\frac{3}{2} &=&\frac{1}{8} \left[ c-5- \sqrt{(1-c)(9-c)} \right]
\ena
one finds
\EQ
\b_G=  \frac{\k+\frac{3}{2}}{\k+1} \b
\EN

For the $N=2$ theories the cosmological term, written as a chiral integral,
is not renormalized, and therefore the physical scale coincides with the
 renormalization scale. Hence
{\em for $N=2$ there is no supergravity dressing of $\b$-functions}.

These results can be verified perturbatively.
We calculate in  the semiclassical limit $c \rightarrow
- \infty$ when
 the  predicted dressing becomes
\bea
 N&=&0:~~~~   \b_G \rightarrow (1+\frac{6}{c}) \b\nonumber\\
N&=&1 ~~~~   \b_G \rightarrow (1+\frac{2}{c}) \b\nonumber\\
N&=&2~~~~   \b_G \rightarrow  \b
\ena
We concentrate on the gravitational dressing of
one-loop $\b$-functions for $\s$-models and to begin with we consider the case
of
bosonic theories. We work in light-cone gauge where the only nonvanishing
component of
the gravitational field is $h_{\mm \mm} \equiv h$ and the induced action
 is
\bea
S_{ind} &=&- \frac{c}{24\pi} \int d^2x~( \pa_{\pp}^2h_{\mm
\mm})\frac{1}{1-\pa_{\mm}^{-1}h_{\mm \mm}\pa_{\pp}}
\pa_{\pp}\pa_{\mm}^{-1}h_{{\mm \mm}} \nonumber\\
&=& -\frac{c}{24\pi}\int d^2x \left[ h\frac{\pa_{\pp}^3}{\pa_{\mm}}h
-h\left(\frac{\pa_{\pp}^2}{\pa_{\mm}}h\right)^2
-\left(h\frac{\pa_{\pp}^2}{\pa_{\mm}}h\right)
\frac{\pa_{\pp}}{\pa_{\mm}}\left(h \frac{\pa_{\pp}^2}{\pa_{\mm}}h
\right) + \cdots \right]
\ena
We are using the conventions of ref. \cite{MGPVN}
with space-time light-cone coordinates denoted by $x^{\pp}$ and
$x^{\mm}$.

We consider a bosonic $\s$-model parametrized by scalar fields
$\phi^i(x)$ and
a target manifold metric $g_{ij}(\phi )$, and described by the action (in the
presence of light-cone induced gravity)
\EQ
S[\phi]
= - \frac{1}{2} \int d^2x~ g_{ij} (\phi )\left( \pa_{\mm} \phi^i - h_{\mm \mm}
\pa_{\pp}\phi^i \right)\pa_{\pp}\phi^j
\EN
We perform a conventional $\b$-function calculation by expanding the action
in normal coordinates
\bea
S=&&S[\phi] - \frac{1}{2} \int d^2x  ~g_{ij} \left( \pa_{\mm} \xi^i - h_{\mm
\mm}
\pa_{\pp}\xi^i  -2 h_{\mm \mm} \pa_{\pp} \phi^i \right)\pa_{\pp}\xi^j
\nonumber\\
&& -\frac{1}{2} \int d^2x R_{ik\ell j} \left(\pa_{\mm}\phi^i -h_{\mm
\mm}\pa_{\pp}\phi^i\right) \pa_{\pp}\phi^j \xi^k \xi^{\ell} \nonumber\\
&&-\frac{1}{3} \int d^2x R_{ik\ell j} \left( \pa_{\mm}\phi^i \pa_{\pp}\xi^j
+\pa_{\pp}\phi^i
\pa_{\mm}\xi^j \right) \xi^k \xi^{\ell} +\cdots
\ena
We note a subtlety:  in a conventional quantum-background splitting of the
covariantized lagrangian $g_{ij} \Del_a \phi^i \Del_a \phi^j \rightarrow
g_{ij}[ \Del_a \phi^i \Del_a \phi^j +2 \Del_a \phi^i \Del_a \xi^j + \Del_a\xi^i
\Del_a \xi^j +\cdots ]$ one would be tempted to drop the middle term because it
is linear in the quantum field $\xi$. However, the covariant derivative
contains the
quantum field $h_{\mm \mm}$; hence the middle term is quadratic in quantum
fields and must be kept.

As it is standard in $\s$-model quantum calculations we refer the $\xi^i$
fields
to tangent space frames, i.e. $\xi^a = e_i^a(\phi) \xi^i$.
The momentum space propagators are (in our light-cone conventions \cite{MGPVN}
$i \pa \rightarrow \frac{1}{2}q$ )
\bea
<\xi ^a\xi ^b> &=& 4 \d ^{ab}\frac{1}{q_{\pp}q_{\mm}} \nonumber\\
<h ~h> &=&- \frac {48\pi}{c} \frac{q_{\mm}}{q_{\pp}^3}
\ena
The one-loop $\b$-function in the absence of gravity is obtained from the
divergence of the tadpole diagram in Fig.1a,  generated by  Wick-contracting
the two quantum fields in the second line of Eq. (17):
\EQ
\G_0 = - \frac{1}{2} \int d^2x R_{ij} \pa_{\pp}\phi^i\pa_{\mm}\phi^j \int
\frac{d^2q}{(2\pi )^2}\frac{4}{q_{\pp}q_{\mm}}
\EN
The tadpole integral, after separating out an IR divergence, leads to the usual
$1/ \epsilon $ UV divergence.

\let\picnaturalsize=N
\def\picsize{5.0in}
\def\picfilename{Fig1.eps}
\ifx\nopictures Y\else{\ifx\epsfloaded Y\else\input epsf \fi
\let\epsfloaded=Y
\centerline{\ifx\picnaturalsize N\epsfxsize \picsize\fi
\epsfbox{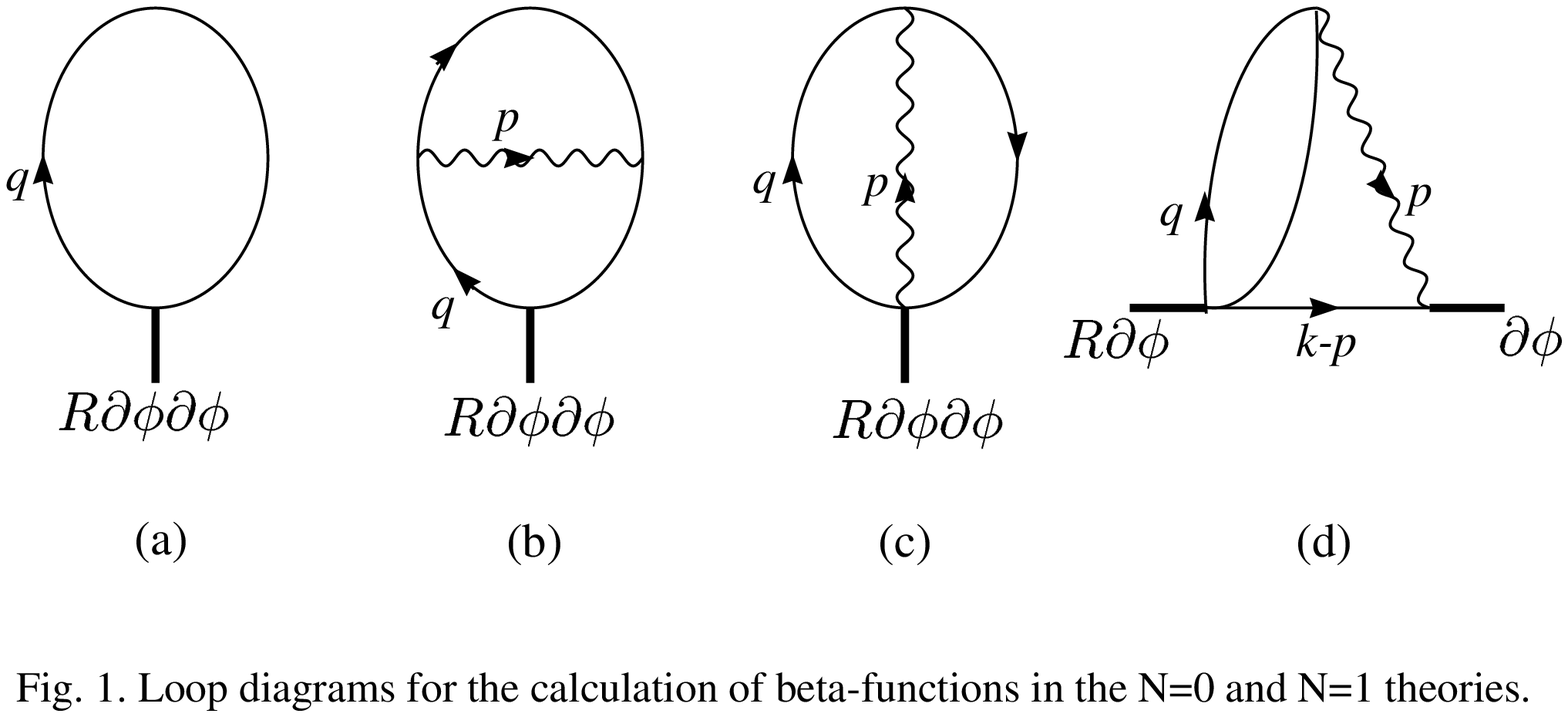}}}\fi

We obtain gravitational corrections to ${\cal O}( 1/c)$ by evaluating the
diagrams
in Fig.1b,c,d. However, it is trivial to verify (and follows from Lorentz
invariance)
that the contribution from the diagram in Fig.1c vanishes. The contribution
from
Fig.1b is given by
\bea
\G_1^{(b)} & &= - \frac{1}{2} \int d^2x R_{ij} \pa_{\pp}\phi^i\pa_{\mm}\phi^j
\nonumber\\
&&\times \int \frac{d^2q d^2p}{(2\pi )^4}  \left(- \frac{48 \pi}{c} \right)
\frac{4}{q^2_{\mm}} \frac{p_{\mm}}{p_{\pp}^3} \frac{(q-p)_{\pp}}{(q-p)_{\mm}}
\ena
We perform the $p$ integral using the methods and the table of integrals of
ref. \cite{MGPVN}, in particular (A.5)
\footnote{Note some misprints: in the right hand side of (A.3)  the exponent
should be
$n-1$ and  the right hand side of  (A.7) should contain a factor $1/p_-$;
the right hand side of (A.9) should be multiplied by
$(-1)^n p_-^{n-1}$.}
\EQ
\int d^2p  \frac{(q-p)_{\pp}p_{\mm}}{(q-p)_{\mm}p_{\pp}^3} =\frac{\pi
q_{\mm}}{2q_{\pp}}
\EN
and we find that $\G_1^{(b)}=-(6/c) \G_0$.

{}From Fig. 1d we obtain the contribution
\bea
\G_1^{(d)} & &=  \frac{1}{3} \int d^2x R_{ij} \pa_{\pp}\phi^i\pa_{\mm}\phi^j
\nonumber\\
&&\times \int \frac{d^2q d^2p}{(2\pi )^4}  \left(- \frac{48 \pi}{c} \right)
\frac{4}{q_{\mm}(p-q)_{\mm}} \frac{p_{\mm}}{p^3_{\pp}} \left[
\frac{(k-p)_{\pp}}
{(k-p)_{\mm}} -\frac{1}{2} \frac{p_{\pp}}{(k-p)_{\mm}} \right]
\ena
which leads to $\G_1^{(d)}= (12/c)\G_0$.

Altogether we obtain therefore
\EQ
\b_0 +\b_1 =\left(1+\frac{6}{c}\right)\b_0
\EN
which agrees in the large $c$ limit with the value obtained from the exact
analysis.

\vspace{0.2in}
We consider now the dressing of $\b$-functions in $N=1$ supersymmetric
$\s$-models
coupled to induced supergravity. In light-cone gauge $ N=1$ supergravity is
described by the superfield $H_-^{\pp}$ \cite{MGRM}. The other geometrical
quantities
are given by
\bea
\Del_+ &=&D_+ \nonumber\\
\Del_- &=& D_- +iH_-^{\pp}\pa_{\pp} -\frac{1}{2}(D_+H_-^{\pp})D_+ +
i(\pa_{\pp}H_-^{\pp})M
\nonumber\\
R&=&iD_+\pa_{\pp}H_-^{\pp} \nonumber\\
E&=&1
\ena
($M$ is a Lorentz generator)
with $\Del_{\pp} = -i (\Del_+)^2$ and $\Del_{\mm} = -i (\Del_-)^2$. The
induced action has the form
\bea
S_{ind} &=& -\frac{c}{4\pi} \int d^2x d^2\th R \frac{\Del_+ \Del_-}{\Box -
\Del^{\a}R\Del_{\a}}R  \nonumber\\
&=&\frac{ic}{16\pi}\int d^2x d^2 \th H_-^{\pp} \frac{\pa_{\pp}^2}{\pa_{\mm}}D_+
D_-H_-^{\pp} +\cdots
\ena
The $\s$-model action and normal coordinate expansion take a form similar to
that of
the bosonic model:
\bea
S &=& - \frac{1}{2} \int d^2x d^2\th g_{ij} \Del_+ \Phi^i \Del_-\Phi^j
\nonumber\\
&& \rightarrow   - \frac{1}{2} \int d^2x d^2\th g_{ij}\left(  \Del_+ \xi^i
\Del_-\xi^j
+ \Del_+ \Phi^i \Del_-\xi^j - \Del_-\Phi^i \Del_+\xi^j \right) \nonumber\\
&& - \frac{1}{2} \int d^2x d^2\th R_{ik\ell j} \Del_+ \Phi^i \Del_-\Phi^j
\xi^k\xi^{\ell}
\nonumber\\
&& -\frac{1}{3} \int d^2x d^2 \theta R_{ik \ell j} \left( \Del_+\Phi^i \Del_-
\xi^j -
\Del_- \Phi^i \Del_+ \xi^j \right) \xi^k \xi^{\ell} +\cdots
\ena
We emphasize again that the terms linear in $\xi$ do lead to quadratic quantum
field contributions and must be kept.

The propagators are
\bea
<\xi^a \xi ^b> &=& 4 \d^{ab} \frac{D_+ D_-}{q_{\pp}q_{\mm}} \nonumber\\
<H_-^{\pp} H_-^{\pp}> &=&  \frac{64 \pi }{c}\frac{D_+D_-}{p_{\pp}^3}
\ena
and the relevant diagrams are again the ones in Fig. 1.

The standard one-loop $\b$-function is obtained from the tadpole diagram in
Fig.1a which, after $D$-algebra, gives
\EQ
\G_0 = -\frac{1}{2} \int d^2x d^2 \th R_{ij}\Del_+\Phi^i \Del_-\Phi^j \int
\frac{d^2q}{(2\pi )^2}
\frac{4}{q_{\pp}q_{\mm}}
\EN
 The dressing is provided by the diagrams in Fig.1b,c,d but again Fig.1c gives
no contribution.
For Fig.1b the relevant vertices are
$-\frac{1}{2} R_{ik \ell j} \Del_+\Phi^i \Del_- \Phi^j \xi^k \xi^{\ell}$ and
 $\frac{1}{2} i H_-^{\pp}D_+\xi^i\pa_{\pp}\xi^i$. This
diagram, because  of three distinct Wick contractions gives
three contributions which, after $D$-algebra, lead to the following result:
\bea
\G_1^{(b)} &=& \frac{1}{2} \int d^2x d^2 \th R_{ij}\Del_+\Phi^i \Del_-\Phi^j
\int
\frac{d^2q d^2p}{(2\pi )^4}
\nonumber\\
 && \times \frac{32\pi}{c} \left[ \frac{(q-p)_{\pp}}{q_{\mm}p_{\pp}^3
(q-p)_{\mm}} + \frac{q_{\pp}}{q_{\mm}p_{\pp}^3(q-p)_{\mm}}
+2\frac{(q-p)_{\pp}}{q_{\mm}p_{\pp}^3(q-p)_{\mm}}
\right]
\nonumber\\
&=& \frac{1}{2c} \int d^2x d^2 \th R_{ij}\Del_+\Phi^i \Del_-\Phi^j \int
\frac{d^2q}{(2\pi )^2}
\frac{8}{q_{\pp}q_{\mm}} \nonumber\\
&=&  -\frac{2}{c}~ \G_0
\ena

For the diagram in Fig. 1d the relevant vertices are $\frac{1}{3} R_{ik\ell j}
\Del_-\Phi^i D_+ \xi^j \xi^k \xi^{\ell}$, $-\frac{i}{2} \Del_+\Phi^i \pa_{\pp}
\xi^i H_-^{\pp}$,
and $-\frac{i}{2} D_+\xi^i \pa_{\pp}\xi^i H_-^{\pp}$. The diagram, after doing
the $D$-algebra
and keeping only divergent contributions gives
\bea
&&\frac{1}{4} \int d^2x d^2 \theta R_{ij} \Del_+ \Phi^i \Del_- \Phi^j \int
\frac{d^2q d^2p}{(2\pi)^4}
\left( \frac{-128 \pi}{c} \right) \frac{1}{q_{\mm}p^2_{\pp}(p-q)_{\mm}}
\nonumber\\
&&=\frac{4}{c} \G_0
\ena
Therefore
\EQ
\b_0+\b_1= \left( 1+\frac{2}{c} \right) \b_0
\EN
which agrees with the exact result in the  $c \rightarrow -\infty$  limit.

Finally, we consider the $N=2$ case.
The $N=2$ $\s$-model is described as usual by a K\"{a}hler potential and,
including the coupling to supergravity, the action takes the form
\EQ
S= \int d^2x d^4\th~E^{-1} K( e^{iH\cdot\pa } \Phi, e^{-iH\cdot\pa}
\bar{\Phi})
\EN
The induced supergravity action is
\EQ
S_{ind}=\frac{c}{2\pi}\int d^2x d^4 \th R \frac{1}{\Box +\cdots} \bar{R}
\EN
where the $\cdots$ indicate curvature dependent terms.
At the linearized level we have the explicit expressions \cite{GZ}
\footnote{The general solution of the constraints of $N=2$
supergravity is given in Ref.\cite{MGMW}.}
\bea
E^{-1}&=& 1-[\bar{D}_+,D_+]H_{\mm} -[\bar{D}_-,D_-]H_{\pp}\nonumber\\
R&=& 4\bar{D}_+\bar{D}_-[\bar{\s}+D_+\bar{D}_+ H_{\mm}+D_-\bar{D}_- H_{\pp}]
\nonumber\\
\bar{R}&=& 4D_+D_-[\s-\bar{D}_+D_+ H_{\mm}-\bar{D}_-D_- H_{\pp}]
\ena
where the vector superfield $H$ and the chiral compensator $\s$ are the
supergravity prepotentials.
The linearized gauge transformations are
\bea
\d H_{\pp}&=& D_+\bar{L}_+-\bar{D}_+ L_+ \nonumber\\
\d H_{\mm}&=& D_-\bar{L}_--\bar{D}_- L_- \nonumber\\
\d\s&=&\bar{D}^2(D_+ L_--D_- L_+) \nonumber\\
\d\bar{\s}&=&D^2(\bar{D}_+\bar{L}_--\bar{D}_- \bar{L}_+)
\ena
We go to a partial light-cone gauge by gauging away $H_{\pp}$ and the
compensator $\s$, $\bar{\s}$. In this gauge the quadratic part of the induced
action takes the form
\EQ
S^{(2)}_{ind}= \frac{2c}{\p}\int d^2x d^4\th H_{\mm}
\frac{\pa_{\pp}}{\pa_{\mm}}
\bar{D}^2 D^2 H_{\mm}
\EN
It still has a residual gauge invariance but, for the present purpose,
instead of using it to gauge away a part of $H_{\mm}$, we prefer to maintain
explicit $N=2$ supersymmetry and do "covariant" gauge fixing. This leads to
propagating ghosts as well; however, since they do not couple to the fields
$\Phi$, they play no role in our calculation.
After gauge-fixing we obtain
\EQ
S^{(2)}_{ind}= \frac{c}{\p}\int d^2x d^4\th~ H_{\mm}\pa_{\pp}^2 H_{\mm}
\EN

 We recall \cite{4loop}  that in the absence of supergravity the computation of
$\b$-functions for $N=2$
 $\s$-models involves a straightforward background-field expansion of the
K\"{a}hler potential $K(\Phi, \bar{\Phi} ) \rightarrow K(\Phi+ \xi , \bar{\Phi}
+ \bar{\xi})$
leading to a quadratic
quantum-field term $K_{\Phi \bar{\Phi}} \xi \bar{\xi}  = \xi \bar{\xi}
+(K_{\Phi \bar{\Phi} }-1)\xi \bar{\xi}$ with the usual chiral propagator
\EQ
<\xi \bar{\xi}> = -\frac{4D^2 \Db^2}{q_{\pp} q_{\mm}}
\EN
As discussed in ref. \cite{4loop}, in order
to compute divergences all the $D$'s and $\Db$'s have to stay inside the loops,
and the summation over the  $K_{\Phi\bar{\Phi}}-1$ vertices leads to an
effective propagator which contains one factor of  the
inverse K\"{a}hler metric $K^{-1}_{\Phi\bar{\Phi}}$.

At one-loop  the $\b$-function is obtained from the UV divergent contribution
to the K\"{a}hler potential
\bea
\G_0 &=& \int d^2x d^4 \th \sum_1^{\infty} \frac{(-1)^n (K_{\Phi  \bar{\Phi}}
-1)^n}{ n}
\int \frac{d^2 q}{(2\pi )^2}
\frac{4}{q_{\pp}q_{\mm}} \nonumber\\
&=& -\int d^2x d^4 \th \ln( K_{\Phi \bar{\Phi}})
  \int \frac{d^2 q}{(2\pi )^2}
\frac{4}{q_{\pp}q_{\mm}}
\ena

\let\picnaturalsize=N
\def\picsize{3.0in}
\def\picfilename{Fig2.eps}
\ifx\nopictures Y\else{\ifx\epsfloaded Y\else\input epsf \fi
\let\epsfloaded=Y
\centerline{\ifx\picnaturalsize N\epsfxsize \picsize\fi
\epsfbox{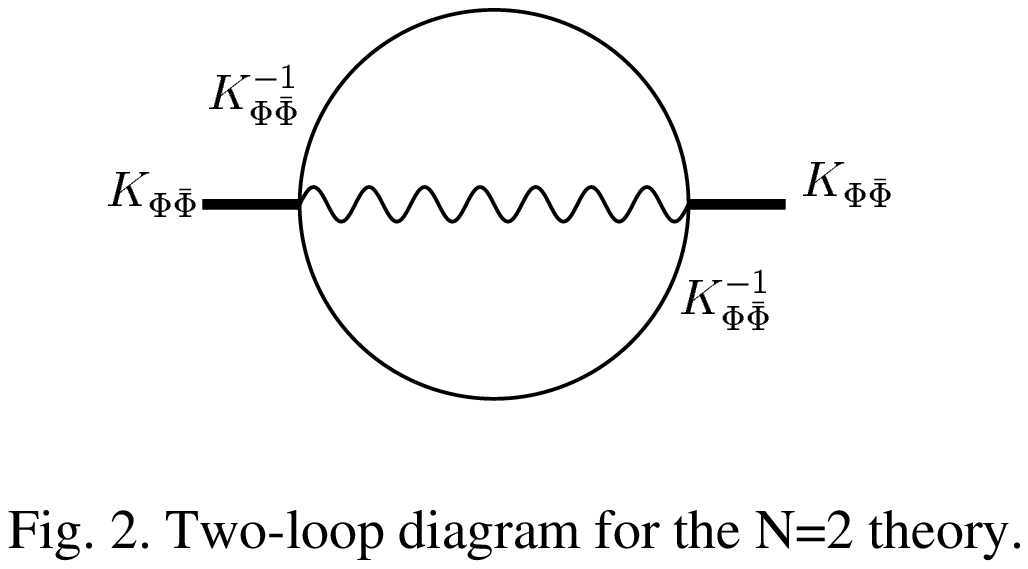}}}\fi

We now consider the gravitational couplings and discuss the two-loop
situation with one supergravity-field exchange.
The  relevant interaction, to lowest order in
$H_{\mm}$ is \cite{GZ}
\EQ
 L_{int} =2H_{\mm} D_+\xi \bar{D}_+ \bar{\xi} K_{\Phi \bar{\Phi}}
\EN
The diagram of interest is given  in Fig. 2 where we have explicitly indicated
the dependence on the background fields at the vertices and in the
matter effective propagators. (The diagram similar to that in Fig. 1d gives no
divergent contributions.) Since, as emphasized above, one obtains a
divergence only when spinor and spacetime derivatives stay in the loops,
irrespective of the details of the loop integrations the dependence on the
$\Phi$ fields cancels completely and no correction to the K\"{a}hler
potential is produced by the supergravity coupling.
Thus in accordance with the general argument presented earlier, there is
no correction to the one-loop $\b$-function in the $N=2$ $\s$-model.

{\bf Acknowledgments}  This research is partially supported by the National
Science Foundation under grant PHY-92-22318.  D. Zanon thanks NSF and MURST,
and M. T. Grisaru
thanks INFN for support.
D. Zanon thanks the Physics Department of
Harvard University and M.T. Grisaru the Physics Department of
Universit\`{a}
di Milano for
hospitality during the period when some of this work was done.

\end{document}